\documentclass[10pt, conference]{FMFP2024}
\usepackage[top=1.50cm, bottom=1.50cm, left=1.884cm, right=1.884cm, includehead, includefoot, heightrounded]{geometry}
\usepackage{fancyhdr} 
\usepackage{graphicx} 
\usepackage[textfont=bf, font = normalsize]{caption} 
\usepackage{floatrow} 
\usepackage{amssymb}
\usepackage{tikz}
\usetikzlibrary{3d}
\usepackage{amsfonts}
\usepackage{amsmath}

\floatstyle{plaintop} 
\restylefloat{table} 
\usepackage{subfigure} 
\renewcommand{\headrulewidth}{0pt}

\begin{document}

\title{\LARGE{\bf Effects of Reynolds number and spatial resolution on the pressure source terms in turbulent boundary layers}}

\author{\textbf{Aditya Agarwal\textsuperscript{1} and Rahul Deshpande\textsuperscript{2}}\\\\
\textsuperscript{1}\small{Department of Mechanical Engineering, Indian Institute of Technology Bombay, Maharashtra 400076, India}\\
\textsuperscript{2} \small{Department of Mechanical Engineering, University of Melbourne, Parkville, VIC 3010, Australia}}

\maketitle
\thispagestyle{fancy} 
\pagestyle{plain} 

\noindent \textbf{ABSTRACT}
\vspace{5pt}

The increase in wall-pressure fluctuations with increasing friction Reynolds number ($Re_{\tau}$) of a turbulent boundary layer (TBL) is well known in the literature.
However, very few studies have investigated the $Re_{\tau}$-variation of the source terms of the pressure fluctuations, which are solely a function of the spatial velocity gradients within the TBL.
This study quantifies the pressure source terms in a zero-pressure gradient TBL by utilizing a published direct numerical simulation (DNS) database \cite{Sillero2013} across 1000 $\lesssim$ $Re_{\tau}$ $\lesssim$ 2000.
It is found that the magnitude of all source terms increases with $Re_{\tau}$ across the entire TBL thickness, with the turbulence-turbulence (non-linear) interaction terms growing faster than the mean-shear (linear) source terms.
Further, we use the simulation database to mimic the scenario of particle image velocimetry (PIV) experiments that are typically spatially under-resolved compared to DNS data.
It is used to quantify the effect of spatial resolution on the accuracy of pressure source terms, which are estimated here for two common PIV scenarios: (i) planar PIV in the streamwise-wall-normal plane, and (ii) stereo-PIV in the spanwise-wall-normal plane of a ZPG TBL.
This exercise reveals significant attenuation of all pressure source terms compared to those estimated from the original DNS, highlighting the challenges of accurately estimating these source terms in a high $Re_{\tau}$ PIV experiment.
\vspace{5pt}
\section{{\textbf{INTRODUCTION}}}
\vspace{5pt}

Boundary layers develop when a fluid flows over a surface, resulting in the generation of a steep mean velocity gradient owing to the no-slip boundary condition at the surface (\emph{i.e.}, wall).
At sufficiently high Reynolds numbers, the mean shear produces turbulent kinetic energy that is significant enough to maintain a chaotic and random fluid motion within the shear layer, commonly known as a turbulent boundary layer (TBL; \cite{Tennekes1972}).
TBLs are commonly noted in nature (such as atmospheric and benthic boundary layers) as well as in various engineering applications related to the aerospace and automotive industries.
Turbulence is responsible for the significant increase in skin-friction drag, heat transfer and flow-induced noise generated at the wall (compared to a laminar flow), all of which influence the efficiency and performance of various engineering systems. 
This study limits itself to the fundamental investigation of the wall-pressure fluctuations ($p_w$), which are well known to influence the structural integrity and flow-induced noise emitted by aircraft and submarine surfaces \cite{Chang1999,Deshpande2024}.

Most past studies \cite{Chang1999, panton2017, tsuji2007} have investigated $p_w$ in relatively low friction Reynolds number range, $Re_{\tau}$ $\lesssim$ $\mathcal{O}$($10^3$), where $p_w$ is predominantly associated with the viscous-scaled (small) near-wall cycle.
Here, $Re_\tau$ is defined as $U_\tau \delta / \nu$, where $U_\tau$ is the mean friction velocity, $\delta$ is the boundary layer thickness, and $\nu$ is the kinematic viscosity. 
However, recent research \cite{Deshpande2024} has demonstrated that $p_w$ increases significantly across $O(10^3) \lesssim Re_\tau \lesssim O(10^6)$, which is particularly relevant for aircraft and submarine operating conditions. 
This increase in $Re_\tau$ is associated with the energization and broadening of the hierarchy of inertia-dominated, large-scale structures/eddies in the outer region of wall-bounded flows. These findings suggest that the effectiveness of $p_w$-attenuating strategies, in high $Re_{\tau}$ flows, would require a thorough investigation of these pressure sources and their coupling with $p_w$.
Understanding and quantifying the sources of wall-pressure fluctuations would also assist with development of $p_w$-predictive models for high $Re_{\tau}$ TBLs, which can inspire improvement in design and performance of various engineering applications \cite{deKat2012}.

To establish the theoretical link between pressure and its sources in the flow field, we begin with the non-dimensionalized Navier-Stokes equation for an incompressible fluid, which are written in standard tensor notation as:
\begin{equation}
\label{eq1}
\frac{\partial \tilde{U}_i}{\partial t} + \frac{\partial}{\partial x_j} (\tilde{U}_i \tilde{U}_j) = -\frac{\partial \tilde{P}}{\partial x_i} + \frac{1}{Re} \frac{\partial^2 \tilde{U}_i}{\partial x_j \partial x_j},
\end{equation}
where \( Re \) is the Reynolds number, \( \tilde{U}_i \) are the instantaneous velocity components, and \( \tilde{P} \) is the instantaneous pressure, with all quantities non-dimensionalized by conventional reference variables.
Per Reynolds decomposition, $\tilde{U}_i$ = $U_i$ + $u_i$ and $\tilde{P}$ = $P$ + $p$, with ($U_i$, $P$) and ($u_i$, $p$) respectively used to denote the mean and fluctuating terms in this manuscript.
$x_i$ represents the streamwise ($x$), wall-normal ($y$) or spanwise ($z$) directions, with $u_i$ denoting velocity fluctuations ($u$, $v$, $w$) along these three directions, respectively.

On taking the divergence of the momentum equation in (\ref{eq1}), implementing the Reynolds decomposition and then subtracting the mean pressure, we obtain the Poisson equation for the fluctuating pressure, which is expressed as \cite{Chang1999}:
\begin{equation}
\label{eq2}
\frac{\partial^2 p}{\partial x_i \partial x_i} = -\left( 2 \frac{\partial {U}_i}{\partial x_j} \frac{\partial {u}_j}{\partial x_i} + {T}^{TT} \right).
\end{equation}
Here, $2 ({\partial {U}_i}/{\partial x_j})({\partial {u}_j}/{\partial x_i}$) is known as the mean-shear ($T^{MS}$) source term for pressure, while $T^{TT}$ is essentially the sum of all the turbulence-turbulence (TT) source terms. 
The individual TT source terms are defined following \cite{Chang1999}:
\begin{equation}
\label{eq3}
T_{ij}^{TT} = \frac{\partial^2}{\partial x_j \partial x_i} ({u}_i {u}_j - \overline{{u}_i {u}_j}).
\end{equation}
The total TT term is defined by the sum of these individual terms:
\begin{equation}
\label{eq4}
T^{TT} = \sum_{i=1}^3 \sum_{j=1}^3 T_{ij}^{TT}.
\end{equation}
The individual \( T_{ij}^{TT} \) terms can be simplified mathematically by implementing continuity, which yields:
\begin{equation}
\label{eq5}
T_{ij}^{TT} = \frac{\partial {u}_i}{\partial x_j} \frac{\partial {u}_j}{\partial x_i} - \frac{\partial^2}{\partial x_i \partial x_j} (\overline{{u}_i {u}_j}).
\end{equation}

Chang et al. \cite{Chang1999} and Ma et al. \cite{ma2021} have previously analyzed the pressure source terms in a turbulent channel flow, which is characterized by a one-dimensional flow in the mean. 
However, in this study we are focusing on zero pressure gradient turbulent boundary layers (ZPG TBL), which are two-dimensional in the mean. 
This distinguishes the relevant terms for analysis in equations (\ref{eq2})-(\ref{eq5}), between ZPG TBLs and a turbulent channel flow. 
For instance, the $T^{MS}$ term needs to be adapted for the case of a ZPG TBL to reflect the gradients of mean statistics along both the streamwise (${\partial}/{\partial}x$) and wall-normal directions (${\partial}/{\partial}y$), {\emph{i.e.}}:
\begin{equation}
\label{eq6}
T^{MS} = 2 \frac{\partial U}{\partial y} \frac{\partial {v}}{\partial x} +2 \frac{\partial V}{\partial x} \frac{\partial {u}}{\partial y}.
\end{equation}
Similarly, the six TT source terms can be expanded from equation (\ref{eq3}) for a ZPG TBL as follows:
\begin{equation}
\label{eq12}
\begin{aligned}
T_{xx}^{TT} &= \left( \frac{\partial {u}}{\partial x} \right)^2 - \frac{\partial^2 \overline{{u}^2}}{\partial x^2},\\
\quad T_{xy}^{TT} = T_{yx}^{TT} &= \frac{\partial {u}}{\partial y} \frac{\partial {v}}{\partial x}  - \frac{\partial^2 \overline{{uv}}}{\partial xy},\\
T_{xz}^{TT} = T_{zx}^{TT} &= \frac{\partial {u}}{\partial z} \frac{\partial {w}}{\partial x},\\
T_{yy}^{TT} &= \left( \frac{\partial {v}}{\partial y} \right)^2 - \frac{\partial^2 \overline{{v}^2}}{\partial y^2}\\
\quad T_{yz}^{TT} = T_{zy}^{TT} &= \frac{\partial {v}}{\partial z} \frac{\partial {w}}{\partial y}, \; \textrm{and}\\
T_{zz}^{TT} &= \left( \frac{\partial {w}}{\partial z} \right)^2.
\end{aligned}
\end{equation}
The total pressure source term is defined as:
\begin{equation}
\label{eq13}
T^{tot} = T^{MS} + T^{TT}.
\end{equation}

Equations (\ref{eq2})-(\ref{eq13}) summarize the relationship between pressure and velocity fluctuations in turbulent boundary layers.
It is evident that all the pressure source terms (\emph{i.e.}, $T^{MS}$ and $T^{TT}_{ij}$) are functions of space and not of time. 
To understand how these source terms contribute to pressure fluctuations ($p$), readers may refer to the Green’s function solution given in Chang et al. \cite{Chang1999}, analysis of which is beyond the scope of the present study. 
The source terms and the Green's function are connected via a convolution integral solved across the flow domain, suggesting that the pressure-fluctuations at the wall, \emph{i.e.} $p_w$ = $p$($y$ = 0), are influenced by the profiles of the source terms across $0$ $\lesssim$ $y$ $\le$ $\delta$. 
This forms the primary motivation of the present study, wherein we focus our attention on the wall-normal profiles of the pressure source terms (both $T^{MS}$ and $T^{TT}_{ij}$) computed from a ZPG TBL DNS database for 1000 $\lesssim$ $Re_{\tau}$ $\lesssim$ 2000.
It is worth noting that the present investigation is an order of magnitude higher than that conducted by Chang et al. \cite{Chang1999} for a turbulent channel flow, providing new insights into the $Re_{\tau}$-variation of the pressure source terms.

\begin{figure*}
\centering
\includegraphics[width=0.8\textwidth]{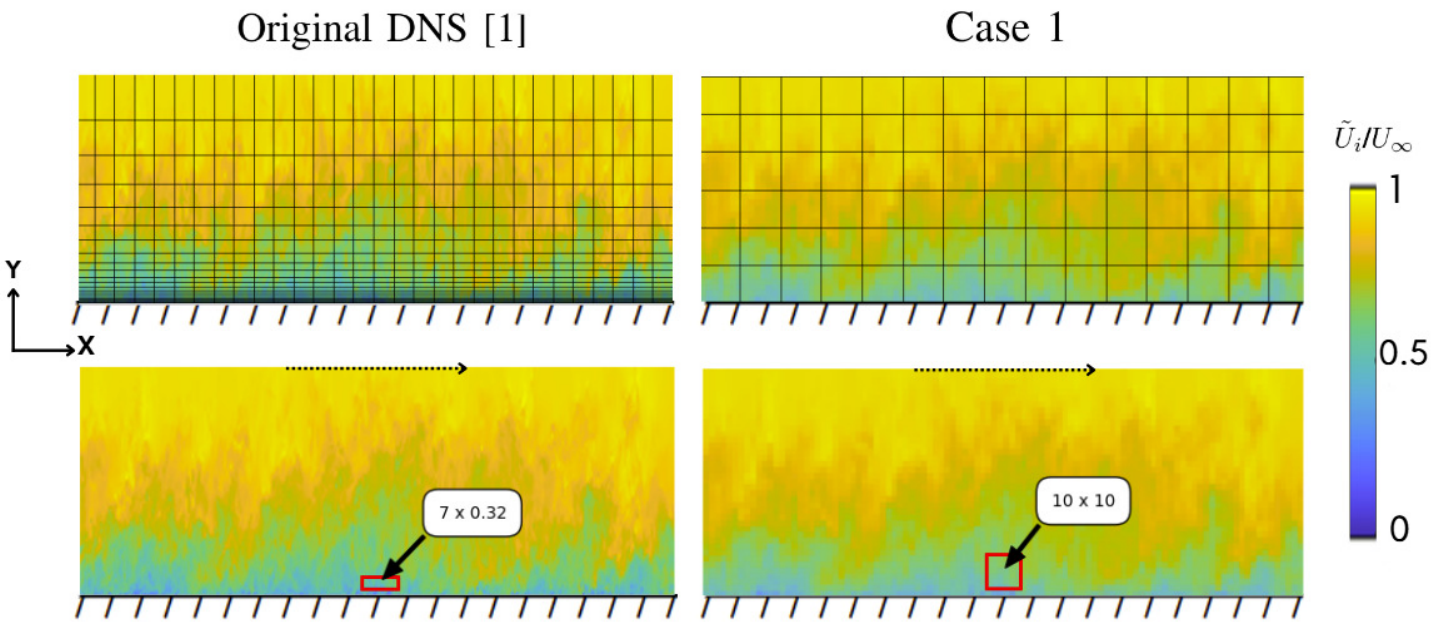}
\vspace{-10pt}
\caption{Comparison of instantaneous streamwise velocity (\( \tilde{U}_i \)/$U_{\infty}$) in an $x$-$y$ plane of a ZPG TBL for different spatial resolutions. 
The left column shows data from the original DNS \cite{Sillero2013} with resolution ${\Delta}x^+$ $\times$ ${\Delta}{y^+_{min}}$ $\times$ ${\Delta}z^+$ $\approx$ 7.0 × 0.32 × 4.07, while the right column depicts the same plane for an under-resolved grid of 10 × 10 × 22. 
The top row illustrates the grid overlay for both resolutions while the bottom row is used to bring out differences in flow features captured by the two grid resolutions.
The wall is at $y$ = 0 and the mean flow direction is from left to right.}
\label{figure1}
\end{figure*} 

The Reynolds numbers realized by DNS in today's date ($Re_{\tau}$ $\sim$ $\mathcal{O}$($10^3$)), however, are significantly lower than the $Re_{\tau}$ associated with several practical applications such as aircrafts and submarines ($\mathcal{O}$($10^4$) $\lesssim$ $Re_{\tau}$ $\lesssim$ $\mathcal{O}$($10^6$)).
Such high $Re_{\tau}$ can only be accessed via experiments in dedicated large-scale boundary layer facilities, wherein the spatial velocity gradients required for computing the pressure source terms can be estimated by three-dimensional (3-D) particle image velocimetry (PIV).
Over the past few decades, PIV \cite{raffel2018} has emerged as one of the key measurement techniques that allows for non-intrusive, detailed measurements of the turbulent flow features over a range of spatial scales.
However, a significant challenge with PIV is its inability to resolve the entire hierarchy of energetic scales in a high $Re_{\tau}$ TBL \cite{Deshpande2023,Lee2016}, spanning from the smallest viscous scales ($\sim$ $\nu$/$U_{\tau}$) to the large inertia-dominated scales ($\delta$).
As demonstrated by Lee et al. \cite{Lee2016}, PIV measurements are prone to spatial attenuation owing to the finite size of the `interrogation volumes', leading to spatial averaging effects that diminish the accuracy of the captured turbulence statistics.
As a consequence, the small-scale turbulent motions are often under-resolved in PIV measurements of a high-$Re_{\tau}$ TBL, resulting in an underestimation of the turbulence intensities \cite{Lee2016}. 
Drawing inspiration from the study of Lee et al. \cite{Lee2016}, here we investigate the effects of spatial resolution on the pressure source terms for the first time.
For this, we use the ZPG TBL DNS data to mimic the scenario of an under-resolved PIV experiment, by following the same strategy adapted by Lee et al. \cite{Lee2016}. 
This involves `low-pass' filtering the fully-resolved DNS data, to the resolution levels expected in a typical PIV experiment.
Pressure source terms estimated from this under-resolved data will be compared against those estimated from the original DNS, to quantify the inaccuracy.
This comparison would be useful for the design and interpretation of future high-$Re_{\tau}$ PIV experiments that are used to estimate the pressure source terms.
We acknowledge here that there are several other plausible sources of error in a typical PIV experiment \cite{raffel2018}, but this investigation limits itself only to those arising from spatial resolution issues.

\section{\textbf{Dataset and Methodology}}
\vspace{5pt}
\subsection{\textbf{Direct numerical simulation database}}
\vspace{5pt}

\begin{table}[t]
\centering
\caption{Grid resolution of the various cases analyzed. Dimensions in bold represent PIV laser sheet thickness}
\begin{tabular}{|c|c|}
\hline
\textbf{Case} & \textbf{Grid Resolution: ${\Delta}x^+$ $\times$ ${\Delta}{y^+_{min}}$ $\times$ ${\Delta}z^+$} \\ \hline
Original & 7.0 $\times$ 0.32 $\times$ 4.07\\ \hline
Case 1 & 10 $\times$ 10 $\times$ \textbf{22} \\ \hline
Case 2 & \textbf{22} $\times$ 10 $\times$ 10 \\ \hline
\end{tabular}
\label{table1}
\end{table}

This study utilizes the publicly available ZPG TBL DNS dataset of Sillero et al. \cite{Sillero2013}, which has been used extensively by past studies investigating spatial resolution issues \cite{Chin2011,Silva2012,Lee2016}.
The dataset comprises of thirteen raw DNS volumes in total, each comprising $\tilde{U}$, $\tilde{V}$, $\tilde{W}$ and $\tilde{P}$ across a very long computational domain, over which the ZPG TBL grows across 900 $\lesssim$ $Re_{\tau}$ $\lesssim$ 2000.
The original resolution of the DNS is ${\Delta}x^+$ $\times$ ${\Delta}{y^+_{min}}$ $\times$ ${\Delta}z^+$ $\approx$ 7.0 $\times$ 0.32 $\times$ 4.07, which has been documented in table \ref{table1}.
Here, ${\Delta}{y^+_{min}}$ corresponds to the smallest wall-normal grid size for the original DNS data, which is noted for the grid element immediately besides the wall.
Detailed descriptions and analyses of the turbulent fluctuations can be found in the published works by Sillero and co-workers \cite{Sillero2013,Sillero2014}, where they have provided comprehensive validations and comparisons with statistics from previously published data.

We extracted two subsets of the full computational domain along the streamwise direction (each nominally $\sim$10$\delta$ long), from each of the thirteen 3-D volumes available to us, such that we obtained thirteen 3-D volumes associated with two different $Re_{\tau}$: 1000 and 2000 (noted at the centre of the respective 3-D volumes).
Limiting the length of the 3-D blocks to 10$\delta$ ensured that the $Re_{\tau}$ did not change significantly across their streamwise extent.
These 3-D volumes were used to compute the wall-normal profiles of the pressure source terms ($T^{MS}$ and $T^{TT}_{ij}$; refer equations (\ref{eq6})$-$(\ref{eq12})) in a ZPG TBL at $Re_{\tau}$ $\approx$ 1000 and 2000.
It permits investigation of the $Re_{\tau}$-variation of the pressure sources terms for the first time (in the limited knowledge of the authors), which will be discussed in $\S$\ref{sec3a}.

\begin{figure}
\centering
\includegraphics[width=0.7\textwidth]{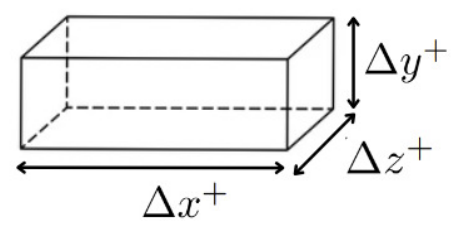}
\caption{Schematic representation of a 3-D (${\Delta}x^+$ $\times$ ${\Delta}{y^+}$ $\times$ ${\Delta}z^+$) grid element to associate with different grid resolutions of the: (i) original DNS, (ii) Case 1 (10 $\times$ 10 $\times$ 22) and (iii) Case 2 (22 $\times$ 10 $\times$ 10), which are documented in table \ref{table1}.}
\label{figure2}
\end{figure}

\subsection{\textbf{Methodology}}
\vspace{5pt}

 \begin{figure*}[t]
\centering
\includegraphics[width=1.0\textwidth]{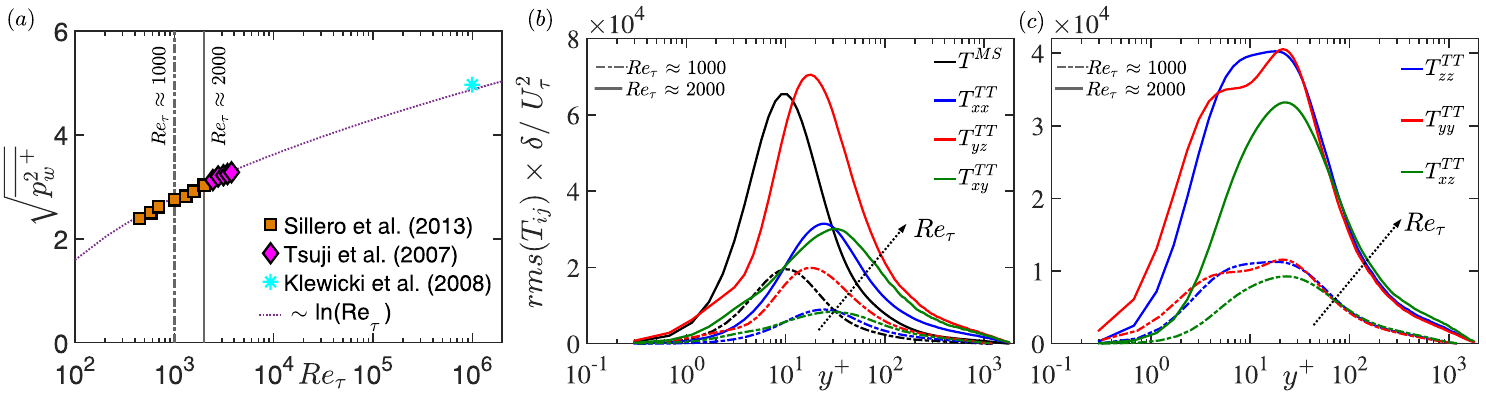}
\caption{(a) $Re_{\tau}$-variation of the root-mean-square (RMS) of $p_w$ documented in the literature by Sillero et al. \cite{Sillero2013}, Tsuji et al. \cite{tsuji2007}, and Klewicki et al. \cite{Klewicki2008}. 
Wall-normal profiles of the RMS values of the pressure source terms: (b) $T^{MS}$, $T^{TT}_{xy}$, $T^{TT}_{xx}$, and $T^{TT}_{yz}$ and 
(c) $T^{TT}_{xz}$, $T^{TT}_{yy}$, and $T^{TT}_{zz}$, for $Re_{\tau} \approx$ 1000 (dashed-dotted) and 2000 (solid lines).
$p_w$ is normalized in viscous units (denoted by superscript $+$), while the source terms are normalized by $\delta$ and $U^2_{\tau}$ \cite{Chang1999}.}
\label{figure3}
\end{figure*}

The DNS dataset was spatially filtered to the target resolutions, documented as Case 1 and Case 2 in table \ref{table1}, by box filtering the velocity fields within the defined interrogation volumes. 
Case 1 mimics the scenario of planar PIV experiment that estimates velocity vectors in an $x$-$y$ plane, having laser sheet thickness along $z$, while Case 2 represents a stereo-PIV experiment in the cross-plane of the flow with laser sheet thickness along $x$.
These dimensions have been chosen based on actual grid sizes noted in past PIV experiments \cite{Lee2016}.
Figure~\ref{figure1} compares the non-dimensionalized instantaneous streamwise velocity ($\tilde{U}_i$/$U_{\infty}$) associated with the grid resolution of the original DNS (left column) and the under-resolved grid corresponding to Case 1 (right column);
the representative grids for the original DNS and `simulated' PIV experiment (Case 1) are overlaid on the top of the fields. 
The instantaneous flow fields shown below highlight the differences in the flow features that are resolved by the two grids.
It is evident that the small-scale features are much better resolved in the original DNS than in Case 1, which is expected.
While figure~\ref{figure1} depicts differences in the 2-D plane, figure \ref{figure2} shows a representative 3-D grid element, the size of which varies from Case 1 to Case 2.

Conventional turbulent statistics such as the normal and shear Reynolds stresses were computed from each of the three cases considered in table \ref{table1} (not shown here for brevity).
Comparison amongst them revealed a significant attenuation of Case 1 and Case 2 statistics relative to the original DNS.
This confirmed that the energy contribution from small-scale eddies is indeed under-resolved for Cases 1 and 2, leading to an underestimation of the Reynolds stresses \cite{Lee2016}. 
The novel aspect in this study is to compute the pressure source terms ($T^{MS}$ and $T^{TT}_{ij}$; equations (\ref{eq6})$-$(\ref{eq12})) from the under-resolved data and compare them against those estimated from the original high-resolution DNS data.
These results are presented and discussed in $\S$\ref{sec3b}.

\section{\textbf{RESULTS AND DISCUSSIONS}}
\label{sec3}
\vspace{5pt}

\subsection{\textbf{Effect of Reynolds number on the pressure source terms}}
\label{sec3a}
\vspace{5pt}

Figure \ref{figure3}(a) compiles the $Re_{\tau}$-variation of the root-mean-square (RMS) of $p_w$ reported from various ZPG TBL datasets in the literature: Sillero et al. \cite{Sillero2013}, Tsuji et al. \cite{tsuji2007}, and Klewicki et al. \cite{Klewicki2008}.
Here, $p_w$ is normalized by viscous length (${\nu}/{U_{\tau}}$) and velocity scales (${U_{\tau}}$), denoted by the superscript $+$.
It is evident that the RMS of $p_w$ increases quasi-logarithmically with increasing $Re_{\tau}$, signifying the importance of investigating the pressure source terms within high-$Re_{\tau}$ TBLs.
At this point, however, we will only analyze the source terms computed from the ZPG TBL DNS database of Sillero et al. \cite{Sillero2013}, at $Re_{\tau}$ $\approx$ 1000 and 2000.

Figures \ref{figure3}(b,c) show the RMS values of the pressure source terms as functions of distance from the wall ($y^+$) for the two $Re_{\tau}$. 
The source terms include the mean-shear term ($T^{MS}$) as well as the turbulence-turbulence interaction terms ($T^{TT}_{ij}$), which are defined in equations (\ref{eq6})$-$(\ref{eq12}) and are normalized by $\delta$ and $U^2_{\tau}$ following Chang et al. \cite{Chang1999}.
It is evident that the magnitude of all the source terms increases with $Re_{\tau}$ across the TBL thickness.
This is consistent with the increase in RMS of $p_w$ (figure \ref{figure3}a) and $p$($y^+$) with $Re_{\tau}$ \cite{panton2017}.
Interestingly, the $y^+$-locations corresponding to the maxima of the pressure source terms remain consistent with increasing $Re_{\tau}$.
For instance, $T^{MS}$ exhibits a maxima at $y^+$ $\sim$ 10 for both $Re_{\tau}$, highlighting the dominant role played by the near-wall mean shear in generating $p_w$.
Similarly, the maxima for all the turbulence-turbulence terms ($T^{TT}_{ij}$) is noted between 20 $\lesssim$ $y^+$ $\lesssim$ 30 for both $Re_{\tau}$.
The comparison between $T^{MS}$ and $T^{TT}_{ij}$ in the outer region of the ZPG TBL is consistent with the previous findings of Kim et al. \cite{kim1989}, who found the $T^{TT}_{ij}$ terms to be much more significant than the $T^{MS}$ term.

A noticeable difference between the two $Re_{\tau}$ cases, however, is the more rapid growth of $T^{TT}_{yz}$ compared to the $T^{MS}$ term in the near-wall region.
This is indicated by the maxima for $T^{TT}_{yz}$ increasing beyond the maxima for $T^{MS}$ at $Re_{\tau}$ $\approx$ 2000, compared to $Re_{\tau}$ $\approx$ 1000.
It suggests the growing importance of the non-linear turbulence-turbulence pressure source terms with increasing $Re_{\tau}$, in both the inner and the outer regions of the TBL \cite{kim1989}.
This hypothesis, however, can be confirmed only after estimating the pressure source terms at much higher $Re_{\tau}$ $\gtrsim$ $\mathcal{O}$($10^4$).
In today's date, such high $Re_{\tau}$ can only be achieved in large-scale experimental facilities, where these pressure source terms can be measured directly via 3-D PIV.
This motivates understanding the effect of spatial resolution on the pressure source terms (refer $\S$\ref{sec3b}), since it can adversely influence the interpretation of high-$Re_{\tau}$ turbulence statistics estimated from PIV \cite{Lee2016}.

\subsection{\textbf{Effect of spatial resolution on the pressure source terms}}
\label{sec3b}
\vspace{5pt}

\begin{figure*}[t]
\centering
\includegraphics[width=1.0\textwidth]{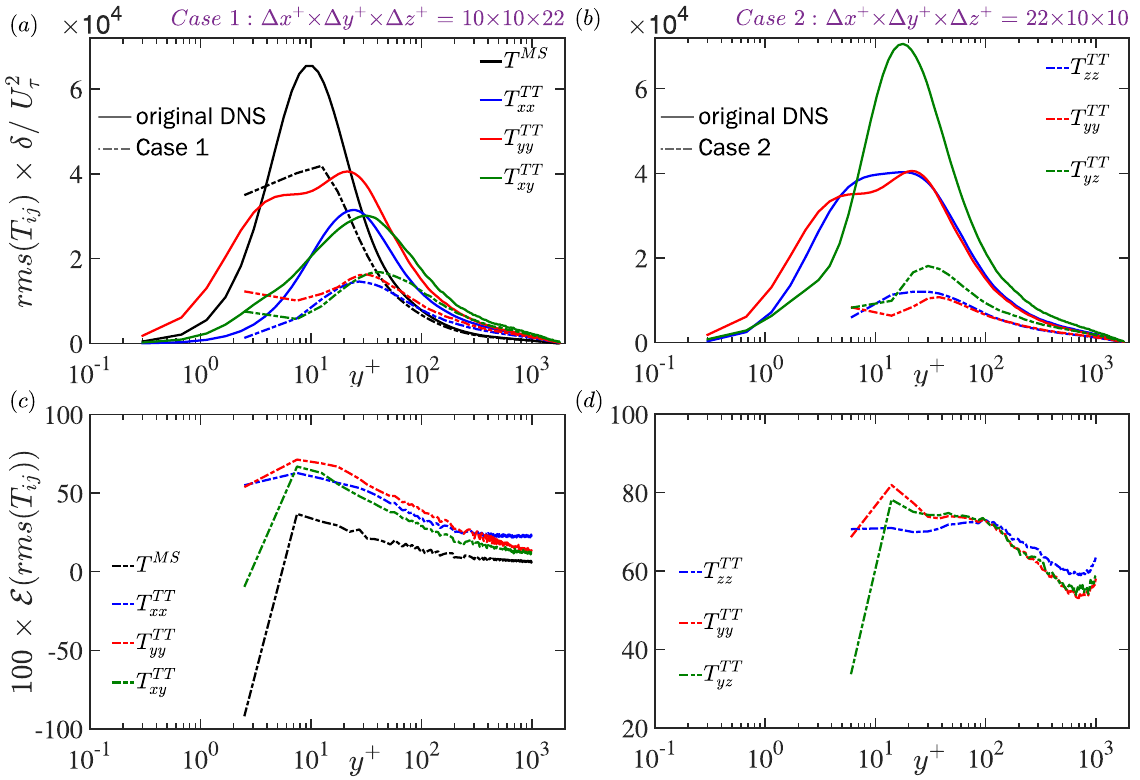}
\caption{(a,b) Comparison of RMS of various pressure source terms computed from the original DNS data (solid lines) and the under-resolved data (dashed-dotted lines) mimicking a PIV experiment corresponding to (a) Case 1 and (b) Case 2.
(c,d) Percentage error in the estimation of RMS of pressure source terms for under-resolved data corresponding to (c) Case 1 and (d) Case 2.}
\label{figure4}
\end{figure*} 

Figure \ref{figure4} showcases the effect of spatial resolution on the RMS of the pressure source terms, by comparing the ones estimated from the original DNS with those estimated from two spatially under-resolved cases. 
Figure \ref{figure4}(a) considers the scenario of a typical planar PIV experiment in an $x-y$ plane, with laser sheet thickness along $z$ (Case 1), while figure \ref{figure4}(b) considers a stereo-PIV experiment in the $y-z$ plane, with laser sheet thickness along $x$ (Case 2).
The fact that both cases resolve data on a single plane means not all seven pressure source terms ($T^{MS}$ and $T^{TT}_{ij}$) can be estimated by either of them.
Following the definitions given in equations (\ref{eq6})$-$(\ref{eq12}), Case 1 can only estimate $T^{MS}$, $T^{TT}_{xx}$, $T^{TT}_{yy}$ and $T^{TT}_{xy}$, while Case 2 can estimate $T^{TT}_{yy}$, $T^{TT}_{zz}$ and $T^{TT}_{yz}$.

In general, the comparison suggests a significant attenuation of all the pressure source terms with decrease in spatial resolution (\emph{i.e.}, increase in ${\Delta}x^+$, ${\Delta}y^+$ or ${\Delta}z^+$). 
Interestingly, however, the $y^+$-locations corresponding to the maxima of the pressure source terms are noted in the similar range for Cases 1 and 2, as noted for the profiles estimated from the original DNS.
To quantify the errors in the RMS of the pressure source terms for both Cases 1 and 2 (relative to original DNS), we plot 100 $\times$ $\mathcal{E}$($rms$($T_{ij}$)) in figures \ref{figure4}(c,d), where $\mathcal{E}$ denotes the error. 
As expected, the errors are maximum in the near-wall region ($y^+ < 10$), where the small-scale energy is predominant in a TBL.
Notwithstanding, the error magnitudes are significant across the entire TBL thickness for both Cases 1 and 2.

Considering the errors for Case 1, $T^{MS}$ has the least error magnitude compared to the other $T^{TT}_{ij}$ terms, suggesting greater sensitivity of the latter to the small-scales.
Since Case 2 only estimates $T^{TT}_{ij}$ terms, their error magnitudes are of the same order amongst themselves.
However, it is worth highlighting here that the error magnitudes for Case 2 are greater than those for Case 1 by $\sim$10\%$-$20\%, indicating the adverse impact of under-resolving along the $x$-direction (Case 2) compared to the $z$-direction (Case 1).
Hence, the orientation and dimensions of the interrogation volumes in PIV experiments play a critical role in the accuracy of the estimated pressure source terms.
We re-emphasize that the uncertainty quantification presented here for the pressure source terms is exclusively associated with the spatial attenuation caused by the interrogation volumes, during typical cross-correlation of particle images. 
The reader is referred to Raffel et al. \cite{raffel2018} for quantifying uncertainty caused by experimental conditions, such as pixel noise (due to imperfections in the imaging process), seeding density, out-of-plane motion, particle image size, etc. 
All these effects interact non-linearly, resulting in additional uncertainty in the turbulence statistics.

The present analysis highlights the challenges of accurately estimating pressure source terms in spatially under-resolved PIV experiments, where the fine-scale turbulence structures are not adequately captured.
These findings align with previous observations of Chin et al. \cite{Chin2011} and Lee et al. \cite{Lee2016}, who demonstrated the adverse impact of poor measurement resolution on the estimation of turbulence intensities.  
Proper consideration of these effects is essential for designing and interpreting future PIV experiments in high-$Re_{\tau}$ TBLs, where the spatial resolution may get worse than that considered for Cases 1 and 2 in the present study.

\section{\textbf{SUMMARY AND CONCLUSIONS}}\label{sec4}
\vspace{5pt}

The present study analyzes the pressure source terms using a previously published ZPG TBL DNS database across 1000 $\lesssim$ $Re_{\tau}$ $\lesssim$ 2000.
It is found that the $Re_{\tau}$-increase in wall-pressure fluctuations \bigg($\sqrt{\overline{p^2_w}^+}$\bigg), which is well documented in the literature, is associated with the increase in RMS of the pressure source terms across the entire TBL thickness.
Despite their increase in magnitude with $Re_{\tau}$, the $y^+$-locations associated with the maxima of the RMS of mean-shear ($T^{MS}$) and turbulence-turbulence ($T^{TT}_{ij}$) remains the same, \emph{i.e.} $y^+_{max}$ $\sim$ 10 for $T^{MS}$ and 20 $\lesssim$ $y^+_{max}$ $\lesssim$ 30 for $T^{TT}_{ij}$.
Present analysis also noted the significantly faster growth of the $T^{TT}_{yz}$ term compared to the $T^{MS}$ term with increasing $Re_{\tau}$, suggesting the growing importance of the turbulence-turbulence interactions in high $Re_{\tau}$ TBLs.
Overall, the analysis discussed in this paper motivates investigation of the pressure source terms in high-$Re_{\tau}$ TBLs, which are more closely related to practical applications such as aircrafts and submarines.

This study also uses the DNS database to examine the impact of PIV spatial resolution issues on the accuracy of pressure source terms.
In general, all the pressure source terms are found to be significantly attenuated, with those associated with the stereo-PIV scenario (Case II; $y-z$ plane) impacted much more significantly than the ones associated with the planar PIV (Case I; $x-y$ plane). 
The $y^+$-locations associated with the maxima of the pressure source terms are, however, preserved despite the loss in resolution.
The present results underscore the necessity of ensuring good spatial resolution for accurate estimation of the pressure source terms for future high $Re_{\tau}$ PIV experiments. 
However, considering some spatial resolution issues are inevitable in experiments, future research should also aim to develop correction schemes \cite{Lee2016} that can assist with recovery of the lost information.

\section{\textbf{ACKNOWLEDGEMENTS}}\label{sec4}
\vspace{5pt}
Rahul Deshpande is grateful for the financial support from the University of Melbourne's Postdoctoral Fellowship.
Both authors gratefully acknowledge Prof. Jimenez and co-workers at the Universidad Politécnica de Madrid, for making their wall turbulence database publicly accessible.

\vspace{0.5cm}
\noindent


\section{\textbf{NOMENCLATURE}}\label{sec4}
\vspace{0.5cm}
\hspace*{-0.9cm}
\begin{tabular}{ccc}
$Re_{\tau}$ & Friction Reynolds number & -- \\

$U_{\tau}$ & Mean friction velocity & [m/s] \\

$\delta$ & Boundary layer thickness & [m] \\

$\nu$ & Kinematic viscosity & [m$^2$/s] \\
$x_i$ & Vector of spatial coordinates & [m] \\
$u_i$ & Velocity fluctuation vector & [m/s] \\
$p$ & Pressure fluctuations & [Pa] \\
\vspace{1pt}
\vspace{1pt}
$T_{ij}^{TT}$ & Turbulence-turbulence source term & [1/s$^2$] \\
\vspace{1pt}
$T_{ij}^{MS}$ & Mean-shear source term & [1/s$^2$] \\
$\sqrt{\overline{p^2_w}}$ & Root-mean-square of wall pressure & [Pa] \\
$+$ & Normalization in viscous units & -- \\
\vspace{1pt}
$\mathcal{E}$ & Error in RMS of pressure source terms & -- \\

\end{tabular}

\bibliographystyle{unsrt}
\bibliography{FMFP2024}

\end{document}